\documentclass[trackchanges,twocolumn]{aastex701}

\usepackage{amsmath}
\usepackage{lipsum}
\usepackage{multirow}
\usepackage{xcolor}
\usepackage{textgreek}
\usepackage[utf8]{inputenc}
\usepackage[english]{babel}
\usepackage{hyperref}
\hypersetup{
    colorlinks=true,
    linkcolor=blue,
    filecolor=blue,      
    urlcolor=blue,
    citecolor=blue,
}
\usepackage{color,colortbl}
\usepackage{tensind}
\tensordelimiter{?}
\DeclareGraphicsExtensions{.bmp,.png,.jpg,.pdf}
\usepackage{verbatim}
\usepackage[normalem]{ulem}
\usepackage{orcidlink}
\usepackage{soul}

\graphicspath{ {./figs/} }

\def\M1450{M_{\rm 1450}}

\def\B21{\citet{becker2021}}
\def\Lya{${\rm Ly} \alpha \ $}
\def\chimp{${\rm {cMpc}}/h\ $}

\newcommand{\absmag}{M_{1450}}
\newcommand{\dlos}{d_{\rm l.o.s.}}
\newcommand{\dsky}{d_{\rm sky}}

\newcommand{\siiv}{Si\,{\sc iv}}

\newcommand{\civ}{C\,{\sc iv}}

\newcommand{\mgii}{Mg\,{\sc ii}}

\begin{document}

\title{The Quasar Proximity Effect as an Alternative Probe of Quasar Pair Distances}

\author[orcid=0000-0002-3211-9642]{Huanqing Chen}
\affiliation{Department of Science, Augustana Campus,
University of Alberta,
Camrose, AB T4V2R3, Canada}
\email{huanqing.chen@ualberta.ca}

\author[orcid=0000-0001-8868-0810]{Camille Avestruz}
\affiliation{Leinweber Institute for Theoretical Physics,
}
\affiliation{Department of Physics,
University of Michigan, Ann Arbor, MI 48109, USA
}
\email{cavestru@umich.edu}

\author{Jakob Wiest}
\affiliation{Department of Physics,
University of Michigan, Ann Arbor, MI 48109, USA
}
\email{jakobw@umich.edu}

\begin{abstract}

Recently discovered quasar pairs at high redshifts ($z\gtrsim$5) are likely precursors to supermassive black hole mergers, providing a promising window to high redshift quasar growth mechanisms.  However, the large uncertainties on their relative distances along the line-of-sight ($\dlos$) limits our ability to characterize quasar pairs. In this study, we explore synthetic quasar proximity zone spectra as an alternative method to constrain the line-of-sight distance of quasar pairs.
We find that for small sky-plane separations ($\dsky\approx 10-100$~pkpc), a simple peak finding algorithm can easily distinguish between scenarios of $\dlos\lesssim1$ pMpc and $\gtrsim1$ pMpc. For cases where the true $\dlos \geq 3$ pMpc, the accuracy of $\dlos$ estimation is $\approx 0.2$~pMpc. Large sky-plane separations of $\dsky=1$~pMpc have larger absolute uncertainties in $\dlos$ estimates, but the method can still easily distinguish between scenarios where $\dlos\lesssim4$~pMpc and $\gtrsim4$~pMpc. $\dlos$ estimates have an uncertainty of $\approx$0.5~pMpc when true $\dlos\gtrsim4$~pMpc. Our proof-of-concept study illustrates the potential use of quasar proximity zones to constrain the 3-dimensional quasar pair configuration, providing an avenue to characterize quasar pairs.
\end{abstract}



\section{Introduction}
\label{sec:intro}
In the past three decades, hundreds of quasars from the Epoch of Reionization (EoR) have been discovered (\citealt{fan1999}; \citealt{fan2001}; \citealt{Fan2023}; also see lists in \citealt{bosman2020zndo}). It has become clear that quasars grow as massive as a billion solar masses within the first billion years after the Big Bang. Such rapid early growth is challenging to explain with the standard Eddington-limited accretion framework for supermassive black hole (SMBH) growth that powers these quasars \citep{inayoshi2020}. Mergers are another major channel for SMBH growth, providing a promising mechanism to help resolve this puzzle \citep{begelman1980,derosa2019}. However, despite being a key ingredient in most semi-analytical models, direct observational evidence of SMBH mergers during the EoR is scarce \citep{zhang2023,jeon2025}.


Yet recently, several close quasar pairs at $z\gtrsim 5$ have been discovered by large ground-based telescopes \citep{djorgovski2003, yue2021,yue2023, matsuoka2024}.
These rare systems are likely precursors to SMBH mergers, providing unique opportunities to understand the SMBH merger mechanism. In particular, two bright quasar pairs with $\mathrm{pkpc}$\footnote{We use ``p'' to distinguish proper kpc (pkpc) and proper Mpc (pMpc) from the comoving units (ckpc and cMpc).} sky-plane separations ($\dsky\approx 1$~pkpc) have been discovered: J2037$-$4537 
($\absmag =-26.42$ and $-25.99$) 
at $z = 5.7$ and J2329$-$0522 
($\absmag =-26$ and $-25$) 
at $z=4.8$ \citep{yue2021,yue2023}.
As likely progenitors of SMBH mergers, we can use these systems to observationally probe the quasar triggering mechanism and the environments of merging quasars.
Apart from these two particularly bright and close quasar pairs, there are other systems that have either larger sky-plane separations or fainter magnitudes.
For example, one of the first bright quasar pairs, SDSS 0338$+$0021 
at $z=5.0$, shows a $\mathrm{Mpc}$ separation \citep{djorgovski2003},
while the recently discovered faint quasar/AGN ($\absmag =-23$) 
pair HSC
J121503.42-014858.7 shows $\dsky\approx$12 $\mathrm{pkpc}$ separation \citep{matsuoka2024}.

A critical limitation to quasar pair characterization is the uncertainty in their true three-dimensional (3D) separation. While the projected sky-plane separation can be accurately measured, the line-of-sight distance ($\dlos$) often carries comparatively large uncertainties.  In turn, the $\dlos$ estimate directly affects the constraint of the quasar-quasar clustering length and merger timescales. 

The most common method for constraining $\dlos$ relies on emission-line redshifts. However, this approach only probes the redshift-space distance and is subject to significant redshift uncertainties, particularly for high-ionization lines that may be broadened or shifted by outflows.
For example, for J2329$-$0522, \citet{yue2023} measured redshifts from multiple emission lines. The redshift differences between the two quasars derived from {\siiv}, {\civ}, and {\mgii} are $0.0082 \pm 0.0073$, $-0.0014 \pm 0.0028$, and $0.011 \pm 0.006$, respectively. At $z=5$, a redshift uncertainty of $\Delta z \sim 0.01$ corresponds to $\sim 6$ physical Mpc (pMpc), or $\sim 36$ comoving Mpc (cMpc)—vastly larger than the projected sky-plane separation of $\sim 10$ pkpc. This large systematic uncertainty in $\dlos$ severely limits our ability to determine whether these quasars are truly physically associated with one another. Therefore, independent methods beyond emission-line redshifts are needed to accurately constrain the three-dimensional separation of these systems. 

The quasar proximity effect \citep{bajtlik1988} offers a promising alternative probe of $\dlos$. Recent studies have used the proximity effect to characterize the local ionizing environment around individual quasars \citep[e.g.,][]{bolton2007, davies2016, eilers2017, chen2021a, Chen2022}. During the EoR, this effect is particularly prominent due to the relatively low ionizing background compared to the quasar's radiation within tens of cMpc. Crucially, quasar radiation impacts the neutral fraction of the intergalactic medium (IGM) in real space rather than redshift space. 
 Therefore, the quasar proximity effect is less susceptible to the finger-of-god effect that plagues emission-line redshift measurements \citep{chen2021b}. When a secondary quasar is present in the foreground of a targeted quasar, we expect a region of enhanced Ly$\alpha$ transmission along the line of sight, providing a direct constraint on their spatial separation.

In this paper, we explore the accuracy of a flux-based detection scheme to constrain $\dlos$ using the proximity effect in quasar pairs as a proof-of-concept for this alternative probe of $\dlos$.  In Section~\ref{sec:methods}, we describe our synthetic sightline data used in our numerical experiments, our detection scheme approach, and the scenario set-ups for our experiments.  In Section~\ref{sec:results}, we present results of our $\dlos$ estimation for various quasar configurations.  Finally, we summarize our conclusions and discussions in Section~\ref{sec:conclusions}.

\section{Methods}\label{sec:methods}

\subsection{Synthetic data}
We generate synthetic quasar spectra from which we develop methods to reconstruct relative l.o.s. distance. We first draw signtlines centered on massive halos from a CROC simulation \citep{gnedin2014,gnedin2014b,gnedin2017}. We choose the top 20 most massive halos in the 40 \chimp box (B40F) at $z=5.2$, drawing 10 sightlines from each halo along random directions.
These sightlines provide gas property information from which we simulate the absorption spectra of the main quasar $Q_A$.

CROC simulations self-consistently calculate ionization of H and He, and the sightlines contain number density of neutral and ionized hydrogen as well as temperature. We calculate the ionization rate from galaxies $\Gamma_{\rm bg}$ assuming ionization equilibrium. If there are quasars presenting,
their radiation greatly reduces the neutral fraction in their immediate surroundings. We calculate the new neutral fraction assuming ionization equilibrium
$$\Gamma n_{\rm HI} = \alpha(T) n_{\rm HII} n_e,$$
where $n_{\rm HI}$, $n_{\rm HII}$ and $n_e$ are the number density of neutral hydrogen, ionized hydrogen and free electron, respectively, $\alpha(T)$ is the recombination rate at temperature $T$, and  $\Gamma$ is the total ionization rate. The total ionization rate is the sum of background radiation (mostly from galaxies), and the radiation from both quasars $Q_A$ and $Q_B$: 
$$\Gamma=\Gamma_{\rm bg}+\Gamma_{A}+\Gamma_B.$$
For quasar radiation $\Gamma_{A}$ and $\Gamma_{B}$, we consider pure geometric attenuation that scales with the distance $r$ from each quasar as $r^{-2}$, which is proven a good approximation in quasar proximity zones \citep{chen2021b}.
With the new ionization fraction, we calculate the \Lya opacity of each cell and generate quasar absorption spectra.
In this paper, we do not consider the uncertainty in quasar continuum \citep{bosman2021}. {We first test quasar spectrum with infinite spectral resolution and zero noise, and in Section \ref{sec:obs}, we discuss the prospects of application to observations.}

\subsection{Scenarios}
We test a few scenarios, varying the magnitudes, l.o.s. distances ($\dlos$) and sky-plane distances ($\dsky$) between the targeted quasar (of which the spectrum is taken) and the surrounding quasar. 
For convenience, we use the following conventions:
\begin{itemize}
    \item Quasar A ($Q_A$) is the quasar we target to observe the spectrum. 
    \item Quasar B ($Q_B$) is the foreground quasar that significantly contributes to the radiation field, impacting the neutral fraction along the line-of-sight towards $Q_A$. 
\end{itemize}

We first set the absolute magitudes at $1450\rm \AA$  of $Q_A$ and $Q_B$ to be $-27$ and $-26$, respectively. These magnitudes are chosen to be similar to the currently known quasar pairs \citep{yue2021,yue2023}. To convert the magnitudes to number rate of ionizing photons, we use quasar templates described in \citet{runnoe2012,lusso2015}, adopting red and blue side of Ly$\alpha$ slopes to be $-0.61$ and $-1.7$ respectively \citep{Chen2022}. We start with a sky-plane distance of $0.01$~pMpc, and vary the $\dlos$ from 0 to 8 pMpc, the uncertainty range for $\dlos$ in recently observed quasar pairs.  We also explore scenarios where we flip the quasar magnitudes (bright quasar in the foreground), and cases with increased sky-plane distance ($\dsky=0.1$~pMpc and 1~pMpc).

\begin{figure*}
    \centering
    \includegraphics[width=0.9\linewidth]{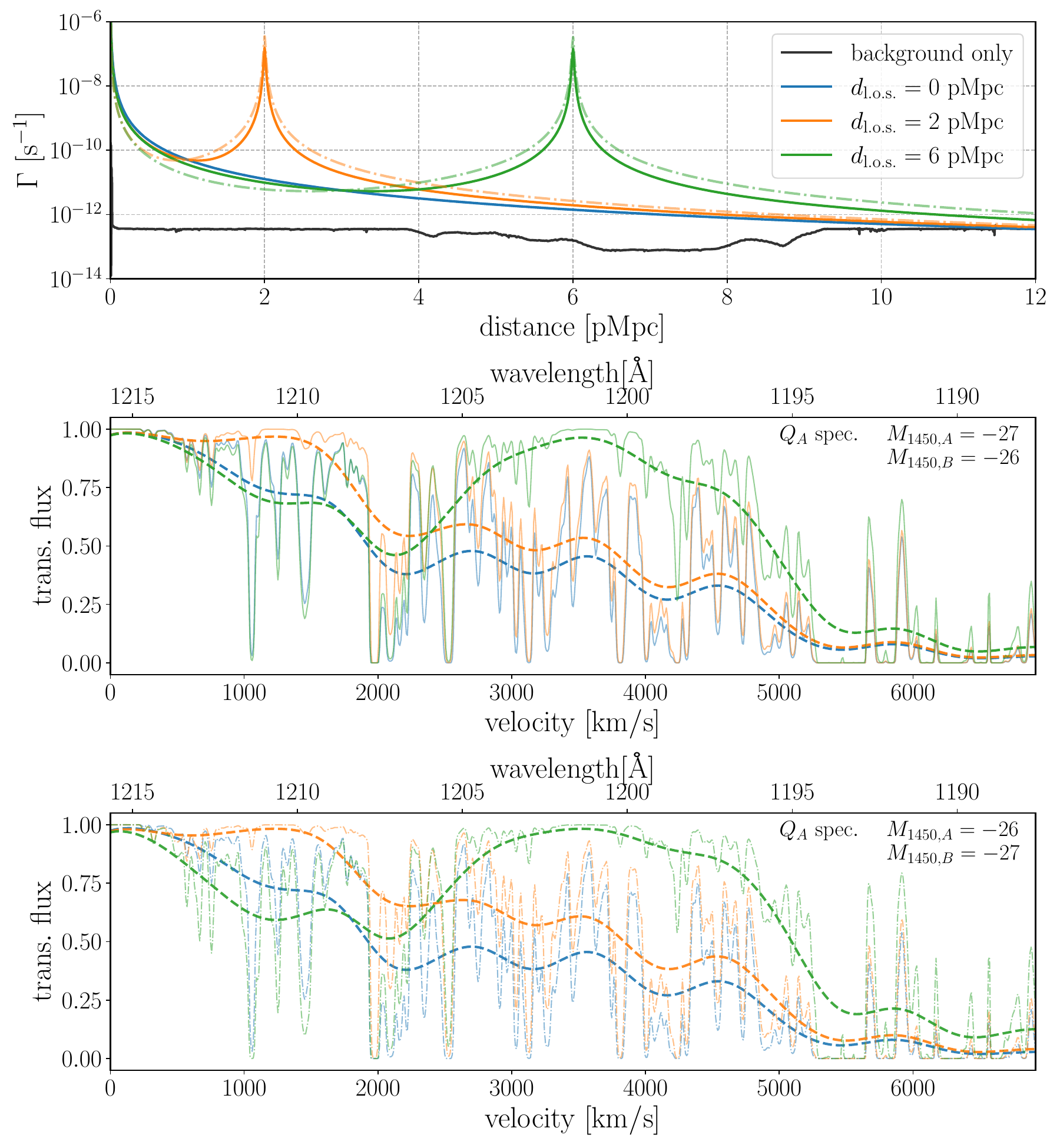}
    \caption{Top: Radiation profile in units of the ionization rate due to both quasars as a function of distance from the target quasar $Q_A$. Solid lines correspond to the scenario where $Q_A$ is the brighter one with $M_{1450}=-27$ and the foreground quasar $Q_B$ with $M_{1450}=-26$. Dash-dotted lines correspond to the inverted scenario where $Q_A$ is the dimmer one with $M_{1450}=-26$ and the foreground quasar $Q_B$ with $M_{1450}=-27$. We fix the sky-plane distance of the quasar pair at 10 pMpc. The three colored lines show the scenarios where the line of sight distance $d_{\rm l.o.s.}$ are 0 (blue), 2 pMpc (orange), and 6 pMpc (green), respectively. Black line corresponds to the contribution from all background galaxies, and the dashed black line the contribution from the target quasar.  Middle: Light solid lines show the corresponding transmitted flux in the above three scenarios with the brighter target quasar, dashed lines trace the Gaussian smoothed spectra.  Bottom:  Light dash-dotted lines show the corresponding transmitted flux for the scenarios with a dimmer target quasar, dashed lines trace the Gaussian smoothed spectra.  The smoothed spectra are very similar for the same $\dlos$, regardless of whether or not the target quasar is the brighter one.}
    \label{fig:gammaflux_smalldsky}
\end{figure*}

For each case, we create synthetic spectra for all the sightlines. In Fig. \ref{fig:gammaflux_smalldsky}, we show the ionization rate (upper panel) and the transmitted spectra for the case of a small plane-of-sky separation, $\dsky=1$~pkpc, comparing a scenario with a brighter target quasar $Q_A$ (solid lines in the middle panel) to a scenario with a brighter foreground quasar $Q_B$ (dashed lines in the lower panel). Different colors show the cases of different $\dlos$. It is clear that the spectra display a distinct bump in transmitted flux corresponds to the corresponding $\dlos$.
In the following sections, we explore the feasibility of predicting $\dlos$ using a simple peak finding scheme on the smoothed spectra.

\subsection{Peak finding on smoothed spectra}

\begin{figure*}
    \centering
    \includegraphics[width=0.39\linewidth]{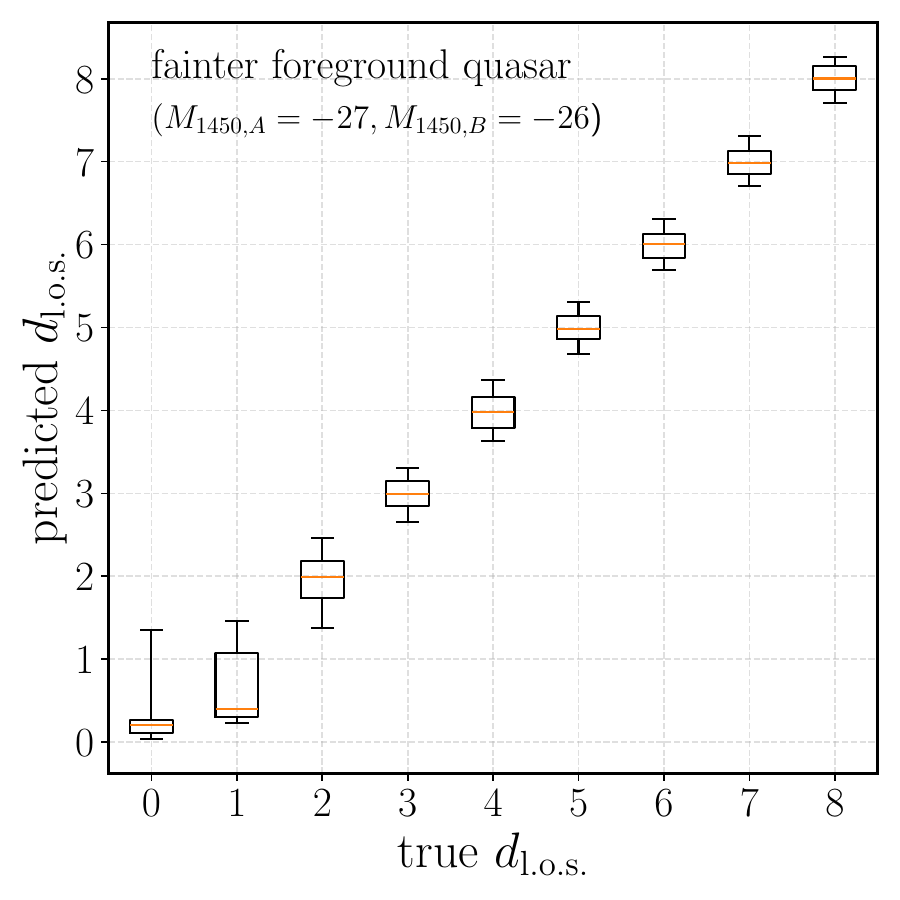}
    \includegraphics[width=0.39\linewidth]{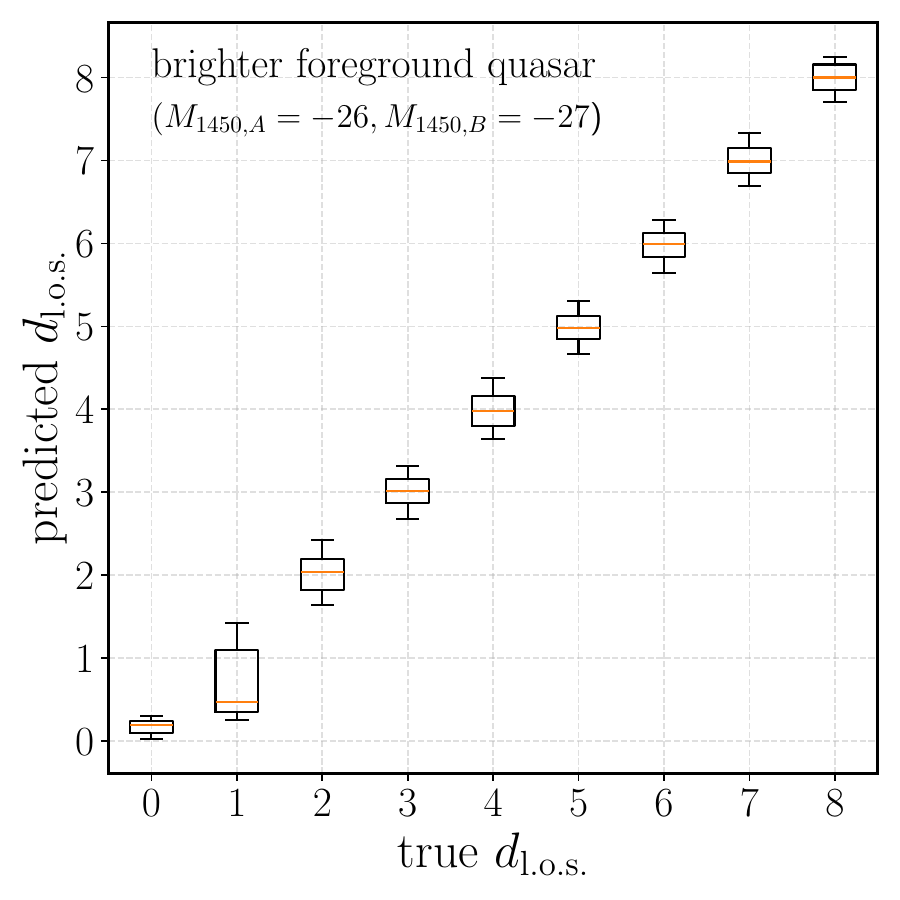}
    \caption{Left: Dependence of predicted $\dlos$ on the true relative distance in the quasar pair configuration case of $d_{\rm sky}=0.01$~pMpc and a brighter background target quasar. The boxes represent 50\% of the $\dlos$ predictions while the bars represent 90\% of the predictions. Peak finding on spectra with a smoothing kernel of 250 km/s is sufficient to measure $\dlos \geq 2~\rm~pMpc$ with consistently tight percentage accuracies. Peak finding can also robustly determine if the foreground quasar's $\dlos$ is smaller than $2~\rm~pMpc$, given the non-overlapping errorbars of $\dlos\geq2~\rm~pMpc$ and $\dlos\leq 1~\rm~pMpc$ cases. Right: Same as left, except for the brighter foreground quasar case. }
    \label{fig:predict_dlos_boxplot1}
\end{figure*}

Density fluctuations strongly impact the transmitted flux. Therefore, it is not straightforward to pinpoint the exact location of the secondary quasar from the spectra. One strategy is to smooth the spectra using a kernel to boost the flux signal corresponding to the true $\dlos$.

For a given quasar, the radiation field of a pencil beam traveling through the center scale with the distance to the quasar $r$ as $\Gamma \propto r^{-2}$. This leads to the scaling relation of neutral hydrogen number density with distance $n_{\rm HI}\propto 1/\Gamma\propto r^2$. Consider a uniform density scenario and the relation between the transmitted flux and the optical depth, this results in a transmitted flux $f$ scaling relation of 
 $f\propto e^{-k r^2},$
 where $k$ is related to the cross-section of the neutral hydrogen and the mean hydrogen number density of the Universe. This gaussian-like shape motivates us to choose a gaussian kernel, which enhance the flux peak at the location of the foreground quasar. For a pencil beam passing by the secondary quasar at a distance, this scaling relationship is also expected when the l.o.s. distance is relatively large compared to the projected sky plane distance. We therefore proceed to smooth the spectra with a gaussian filter. We tested a few choices of the standard deviation of the gaussian kernel to find an optimum kernel size\footnote{{We compared boxcar kernels, which perform worse than gaussian kernels because the elevation of foreground transmission follows a Gaussian-like profile.}}.

\section{Results}\label{sec:results}

Here we describe our results of line-of-sight (l.o.s.) quasar distance ($\dlos$) estimation using a Gaussian smoothing kernel and a simple peak finding algorithm and compare our estimates with the true $\dlos$.

\subsection{Peak finding for brighter target quasar} \label{sec:case1} 
In this subsection, we discuss the scenario where we fix the sky-plane distance to $d_{\rm sky}=0.01$ pMpc and set the target quasar, $Q_A$, to be the brighter quasar with $M_{1450}=-27$.  The foreground quasar $Q_B$ is the fainter one, with $M_{1450}=-26$, in the foreground with varying distance $\dlos>0$.

In the upper panel of Fig. \ref{fig:gammaflux_smalldsky}, we show the radiation profile, $\Gamma$, along the sightline of the target quasar. Solid lines correspond to the scenario of a brighter target quasar ($M_{1450,A}=-27$) and a fainter foreground quasar ($M_{1450,B}=-26$). The blue, orange, and green colors correspond to cases where $Q_B$ are at l.o.s. distance $\dlos=$ 0, 2, 6 pMpc, respectively. Due to the small sky-plane distance, the ionization parameter exhibits a clear peak as high as $\Gamma\approx 10^{-7} \rm s^{-1}$ along the sightline at $\dlos$ of $Q_B$. The additional radiation from $Q_B$ enables further gas ionization, even for relatively dense gas in the cosmic filaments. 

In the middle panel of Fig. \ref{fig:gammaflux_smalldsky}, we show how the additional radiation affects the corresponding transmitted flux in the scenario of the brighter target quasar $Q_A$ and small $\dsky$. Compared with the case where $\dlos=0$, the $\dlos=2$ pMpc case shows an enhanced transmission between distance $\dlos \approx 1.5-3 \rm ~pMpc$, or spectral region corresponding to $H\dlos \approx 1000-2000$ km/s, where $H$ is the Hubble expansion rate at that redshift. In this spectral region, the deep absorption (transmission $<0.1$) in the $\dlos=0$ case is due to gas with densities close to 10 times the cosmic mean ($\Delta_g=10$). While when $Q_B$ is in the foreground, in regions where additional radiation from $Q_B$ leads to ionization rates of $\Gamma\gtrsim10^{-10} ~\rm~s^{-1}$, we see increases in the transmitted flux from near 0 to $>0.75$, or in transmitted fluxes near 1 (e.g. fully transmissive) at spectral regions where there had been some absorption. We see similar results in the $\dlos=6 ~\rm~pMpc$ case, with a notably more significant boost in transmitted flux due to the relatively larger contribution from $Q_B$ than $Q_A$ at $H\dlos \approx 2500-5000$ km/s region.

\begin{figure*}
    \centering
    \includegraphics[width=0.95\linewidth]{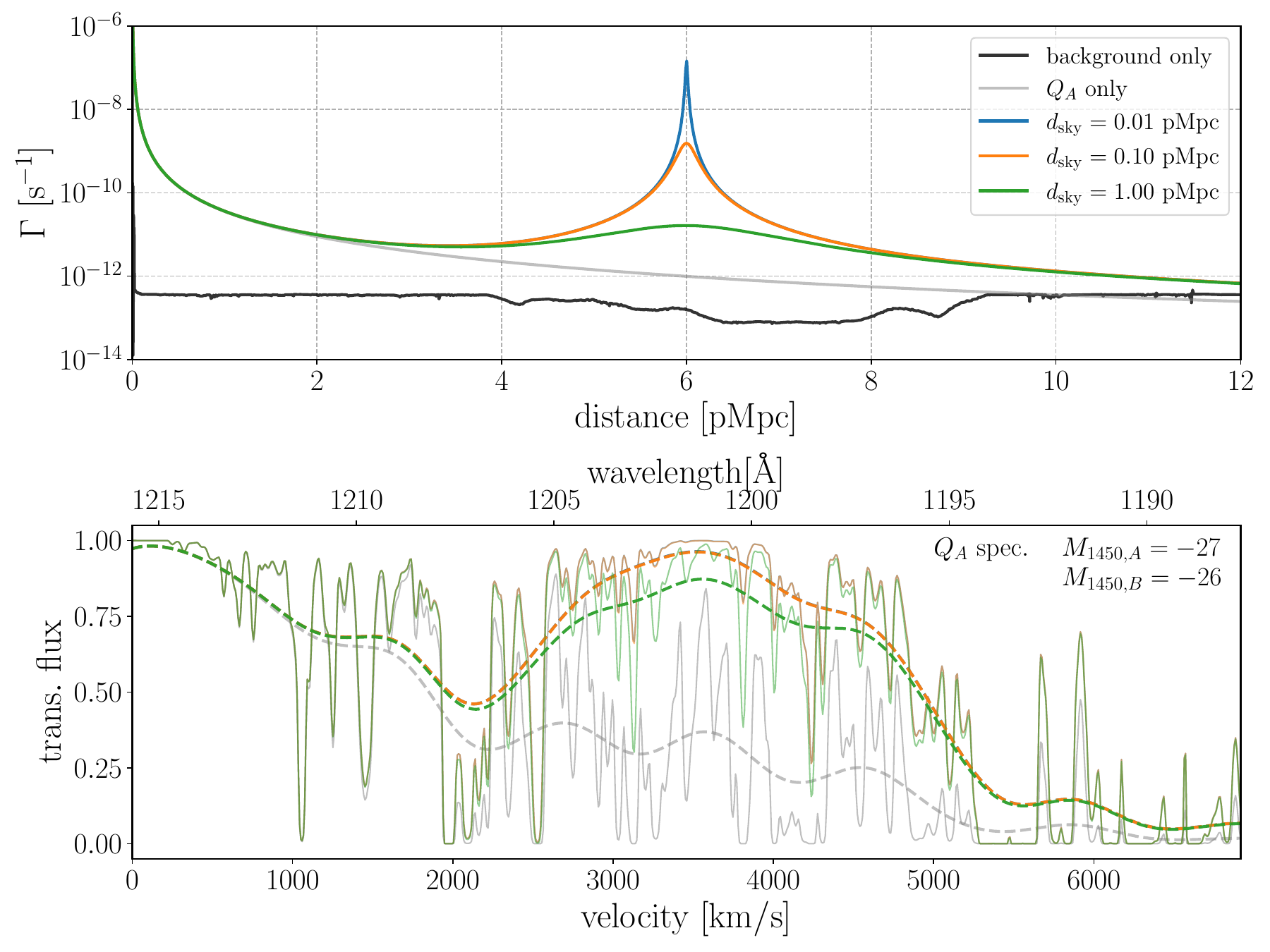}
    \caption{Same figure and quasar pair scenario as Fig.~\ref{fig:gammaflux_smalldsky}, where $Q_A$ is the brighter one with $M_{1450}=-27$.  We fix the line of sight distance to $d_\mathrm{l.o.s.}=6$~pMpc.  The 3 colored lines show the scenarios where where the sky-plane distance $d_\mathrm{sky}$ are 0.01~pMpc (blue), 0.10~pMpc (orange), and 1.0~pMpc (green). The blue and orange lines completely overlaps due to the nearly identical transmitted flux. Grey line is a reference if the secondary quasar $Q_B$ does not exist.}
    \label{fig:vary_dsky}
\end{figure*}

\begin{figure}
    \centering
    \includegraphics[width=0.95\linewidth]{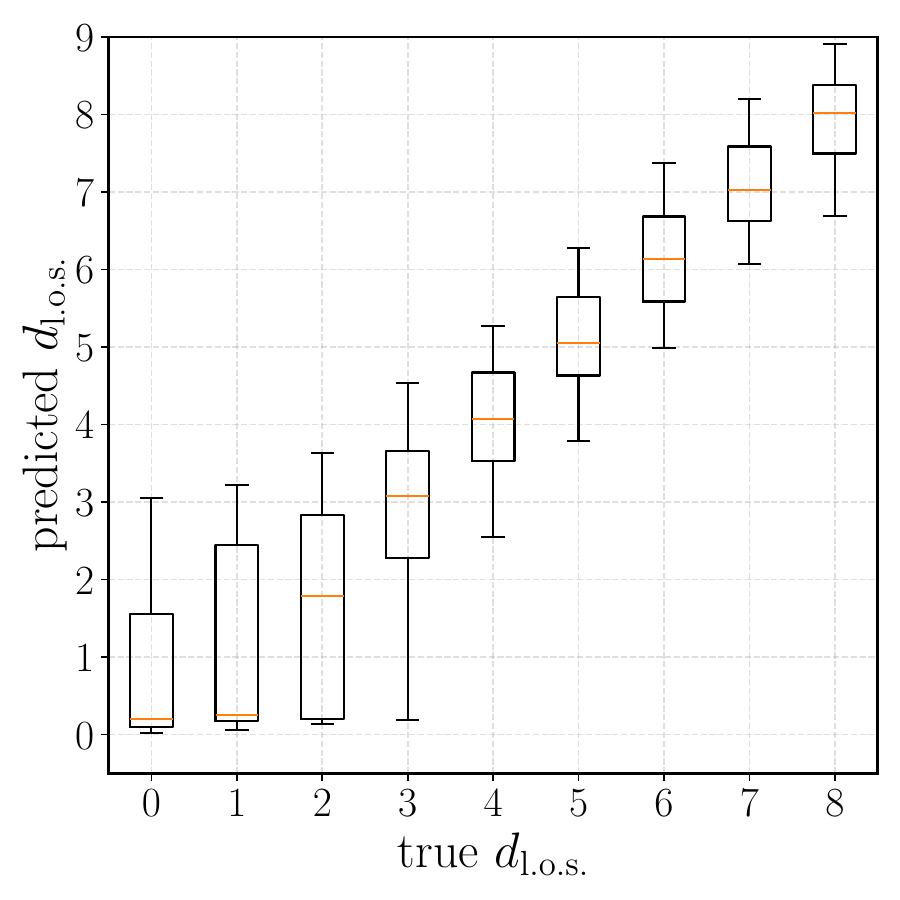}
    \caption{Same as the left panel of Fig.~\ref{fig:predict_dlos_boxplot1}, but for the larger $\dsky=1$~pMpc case.}
    \label{fig:boxplot_case3}
\end{figure}

\begin{figure*}
    \centering
    \includegraphics[width=0.57\linewidth]{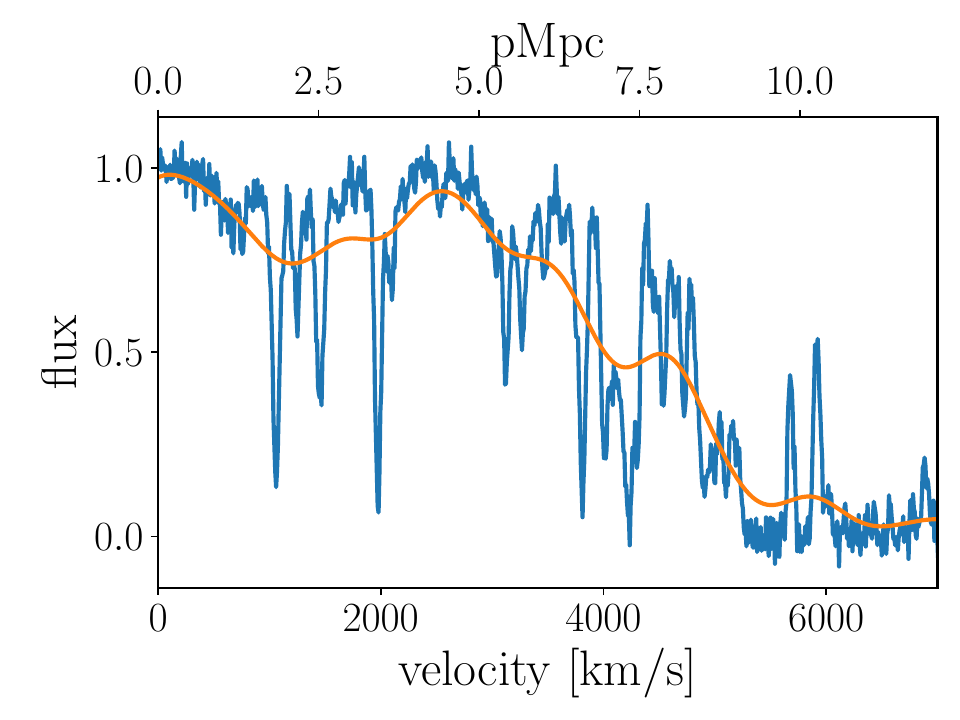}
\includegraphics[width=0.4\linewidth]{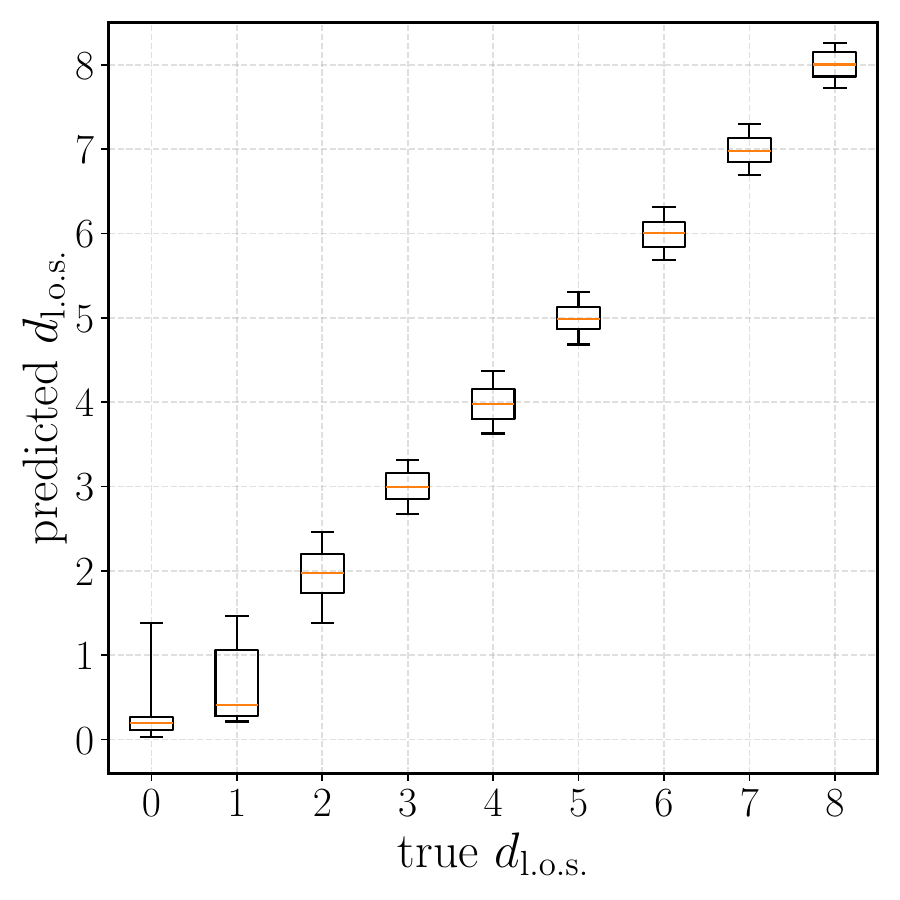}
    \includegraphics[width=0.57\linewidth]{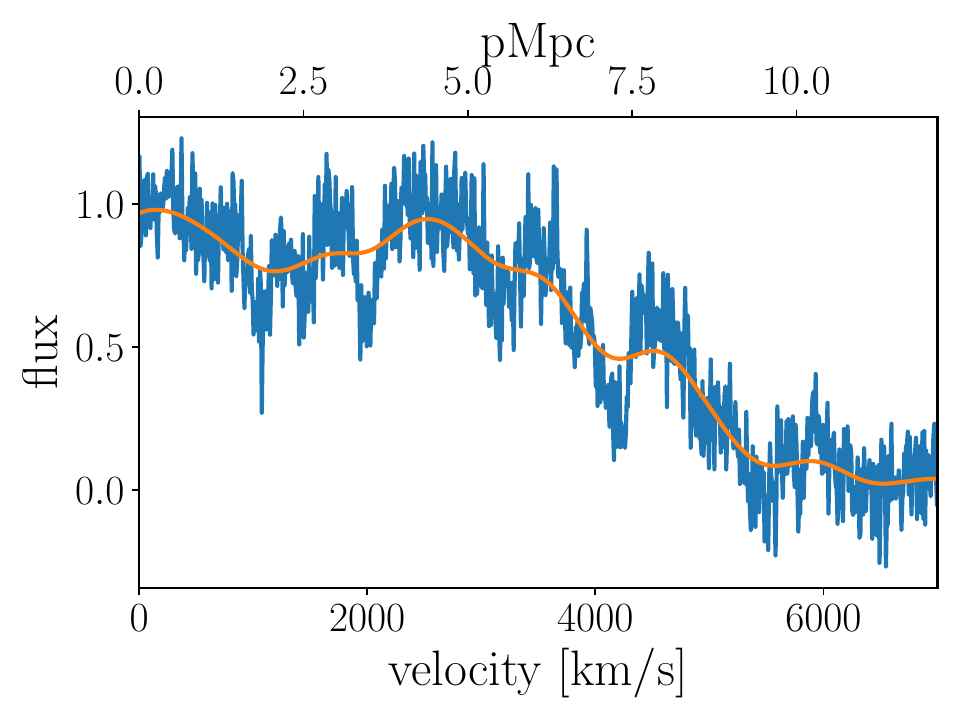}
\includegraphics[width=0.4\linewidth]{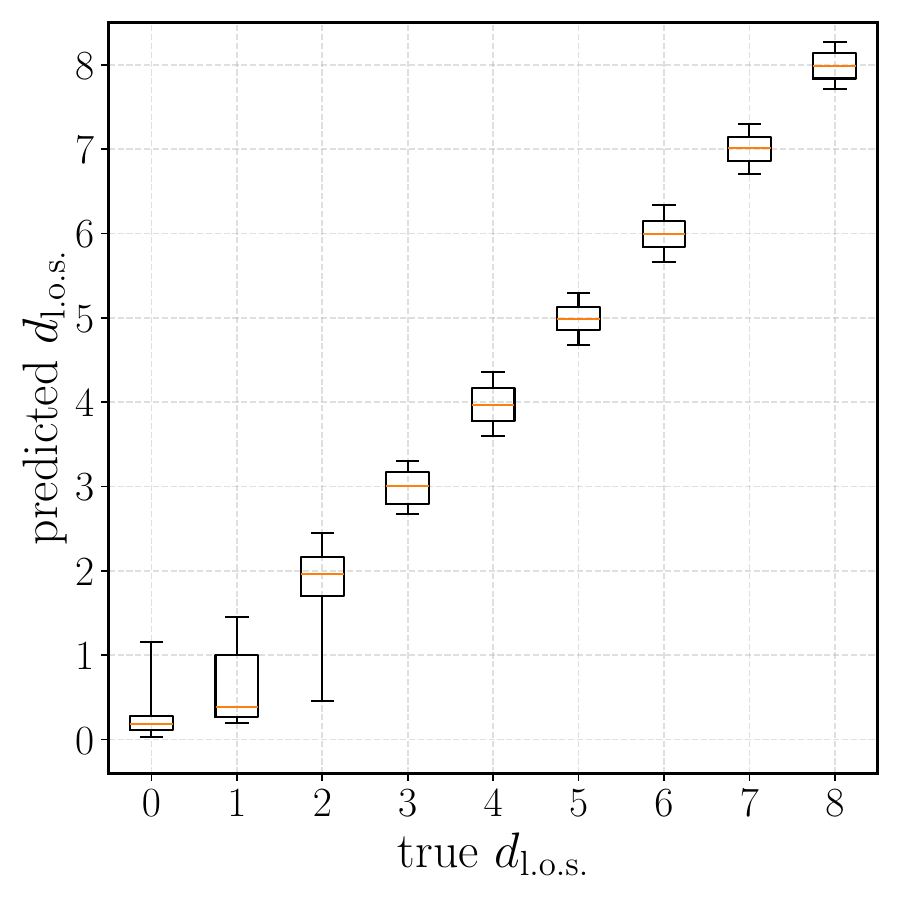}
\caption{Upper panels: the left panel shows a simulated spectra with spectral resolution $R=8000$ and SNR=30 per 10 km/s pixel. Using the same peak finding algorithm, the predicted $\dlos$ vs. true $\dlos$ result is shown on the right. Lower panels: same as the upper panels, except for simulated spectra with spectral resolution $R=2000$ and SNR=10 per 10 km/s pixel.}\label{fig:mockres}
\end{figure*}
Given the strong increase in flux, we can expect that convolving the spectrum with a kernel that corresponds to the size of the secondary $\Gamma$ peak will lead to a smooth spectrum with a flux peak near 1 at a wavelength corresponding to the true $\dlos$ of $Q_B$. We therefore explore a simple peak finding algorithm on a smoothed spectrum with Gaussian kernel of $\sigma=250~\rm~ km/s$\footnote{We have explored different kernel size of $\sigma=125~\rm~ km/s$ and $\sigma=500~\rm~ km/s$ and found that $\sigma=250~\rm~ km/s$ perform the best.} This kernel corresponds to the half-width of the peak in the radiation profile, where $\Gamma$ from the secondary quasar is $\gtrsim 10^{-10}~\rm~s^{-1}$. The dashed lines in the lower panel show such smoothed spectra. We then identify local maxima with the additional condition that the peak must exceed flux limit $f_{\rm lim}=0.8$ to prevent the confusion from ubiquitous transmission spikes.  In the two cases shown, we detect the peak at $\dlos\approx2.1$ pMpc for the orange dashed line and $\dlos\approx6.1$ pMpc for the green dashed lines.  Both estimates are very close to the respective true positions of $Q_B$ $\dlos$, thereby illustrating the potential power of using transmitted flux to constrain the configuration of quasar pair systems.

We show how the predicted $\dlos$ and the percentage accuracy of the prediction changes with the true $\dlos$ in the left panel of Fig.~\ref{fig:predict_dlos_boxplot1}.  Note, we fix the gaussian filter to $\sigma=250$ km/s, which is sufficient to retain reasonably tight percentage accuracies for true relative distances of $\dlos\gtrsim2$~pMpc, or $dz=0.004$.  The percentage accuracy noticeably worsens at smaller relative distances. However, given the non-overlapping errorbars of $\dlos \gtrsim 2$ pMpc and $\dlos \lesssim 1$ pMpc cases, this method can still robustly differentiate these two cases.

\subsection{Peak finding for brighter foreground quasar}

In this subsection, we discuss the small $\dsky=0.01$~pMpc scenario in which we instead set the target quasar, $Q_A$, to be the fainter quasar with $M_{1450}=-26$.  The foreground quasar $Q_B$ is the brighter one, with $M_{1450}=-27$.

We show the radiation profile, $\Gamma$, along the sightline of the target quasar for this scenario in dash-dotted lines in the upper panel of Fig. \ref{fig:gammaflux_smalldsky}. The blue, orange, and green lines show the same $\dlos$ cases as in the previous scenario with a brighter target quasar.  A primary difference in this second scenario of a brighter foreground  quasar is the boost in radiation profile over a larger distance range of the foreground quasar.  When comparing the solid lines to the dash-dotted lines in the top panel, we see that the brighter foreground quasar $Q_B$ broaden the peak in $\Gamma$ at the location of $Q_B$.

In the bottom panel of Fig. \ref{fig:gammaflux_smalldsky}, we show how this radiation from the brighter foreground quasar affects the corresponding transmitted flux, illustrated by the dash-dotted lines in orange (for $\dlos=2$~pMpc) and green (for $\dlos=6$~pMpc. 
The $\dlos=2$ pMpc case shows an enhanced transmission between distance $\dlos=1-3 \rm ~pMpc$, or spectral region of 500-2000 km/s. In the case of $\dlos=6$ pMpc, the transmitted flux profile near the foreground quasar is relatively broader and flatter when compared with the quasar pair scenario with a fainter $Q_B$ (i.e. comparing same color dashed lines in bottom panel with the middle panel).
Similar to the brighter $Q_A$ case, regions with additional radiation contribution from $Q_B$ leads to ionization rates of $\Gamma\gtrsim10^{-10}~\rm s^{-1}$ and transmitted fluxes near 1.

We similarly explore results from our simple peak finding algorithm using a gaussian kernel of $\sigma=250$~km/s. Due to the further enhancement in transmitted flux, when identifying local maxima, we explore additional condition of flux limit $f_{\rm lim}$ between 0.8 and 1.0. We find that setting $f_{\rm lim}$ to 0.9 leads to a more accurate estimate. 
We show how predicted $\dlos$ and percentage accuracy of the prediction changes with the true $\dlos$ in the right panel of 
Fig.~\ref{fig:predict_dlos_boxplot1}. Similar to the previous case, the simple peak finding results in a tight percentage accuracies for true relative distances of $\dlos\gtrsim2$~pMpc, while when $Q_B$ is closer than 1 pMpc it is challenging to pinpoint $\dlos$.

\subsection{Peak finding for larger sky-plane distances}

Here, we test the accuracy of the peak finding prediction for scenarios with larger sky-plane distances,  $d_\mathrm{sky}$.
Fig. \ref{fig:vary_dsky} shows the $\Gamma$ profiles (upper panel) and transmitted flux (lower panel) for the quasar pair configuration where $Q_A$ is the brighter quasar, and $Q_B$ has $d_{\rm l.o.s.}=6$~pMpc.  Each of the blue, orange, and green lines correspond to a different $d_\mathrm{sky}$. 

The blue line, $d_\mathrm{sky}=0.01$~pMpc, is the identical case to the green solid line in Fig.~\ref{fig:gammaflux_smalldsky}.  For increasing values of $d_\mathrm{sky}$, the sharpness of the peak from $Q_B$ in the $\Gamma$ profile (top panel) weakens.  Between the cases of $d_\mathrm{sky}=0.01$~pMpc and $d_\mathrm{sky}=0.1$~pMpc, there is little difference in the $\Gamma$ profile except the peak strength over the narrow $\sim 0.1$ pMpc range.  In fact, the resulting transmitted flux is near-identical. We test the accuracy of the simple peak finding prediction scheme the same as in Sec. \ref{sec:case1} and obtain the near-identical prediction result (left panel of Fig. \ref{fig:predict_dlos_boxplot1}) as well.

However, the $d_\mathrm{sky}=1$~pMpc case results in a notably subdued peak in the $\Gamma$ profile, and less transmitted flux at the corresponding spectral region ($3000-4000$ km/s). We again examine the accuracy of prediction using the peak finding algorithm, this time lower the flux threshold $f_{\rm lim}$. We search the optimal $f_{\rm lim}$ in a step of 0.05, and find that $f_{\rm lim}=0.65$ performs the best. When the flux threshold is low, more transmission spikes associated with cosmic voids are likely to cause ``false positive'' detection of boost radiation field. Therefore, we further explore additional flux peak width requirement. We find that adding the requirement of peak width to be larger than $250$~km/s slightly improve the accuracy. In Fig. \ref{fig:boxplot_case3}, we show the result of the predicted $\dlos$ vs. the true $\dlos$. Compared to the $\dsky=0.01$~pMpc or $0.1$ pMpc case, the accuracy noticeably worsens, especially for $\dlos\lesssim 4$~pMpc.  The decrease in accuracy is expected since the secondary quasar has a relatively lower contribution along the target quasar sightline at small $\dlos$. When the true $\dlos >= 4$ pMpc, the percentage uncertainty improves to $\approx$0.5~pMpc, thanks to the relatively large contribution from the foreground quasar. We have also explored simultaneously changing $f_{\rm lim}$ and the flux peak width requirement, but the improvement in prediction accuracy is not very noticeable.

\subsection{{Prospects of application to observations}} \label{sec:obs}
{Because the radiation impact of the foreground quasar is on large spatial scales, it imposes relatively lenient requirements on spectral resolution to observe the flux boost. As shown in the previous sections, smoothing with a Gaussian kernel of $\sigma=250\rm~km/s$ (${\rm FWHM} = 2.355 \sigma=590~\rm km/s$) results in accurate peak detection on ideal spectra with infinite resolution and signal-to-noise ratio (SNR). This means that as long as the spectral resolution is significantly better than $R=c/\rm FWHM \sim 500 $, the enhancement in transmission can be detected. The left panels of Fig. \ref{fig:mockres} show mock spectra considering limited spectral resolution and SNR, corresponding to a Case 1 scenario where the foreground quasar is at $\dlos=4$ pMpc. The upper panel shows a spectrum with $R=8000$ and SNR = 30 per 10 km/s pixel, e.g., XQR-30 quality \citep{Dodorico2023}, while the lower panel shows a spectra of $R=2000$ and SNR = 10 per 10 km/s pixel. The orange lines show the mock spectra smoothed by a Gaussian kernel of $\sigma=250$ km/s, which clearly peaks near $\dlos=4$ pMpc. The right panels show the predicted $\dlos$ vs. true $\dlos$ results, demonstrating similarly high accuracy (Fig. \ref{fig:predict_dlos_boxplot1}).} 

{The biggest challenge in applying this method to observed quasars is the uncertainty in continuum modeling, which is a common challenge for studies using $z>5$ quasars \citep{bosman2020}. Many algorithms exists, ranging from principal component analysis to more complex machine learning methods \citep{bosman2020,eilers2020,durovcikova2020,greig2024}, and the uncertainty in the quasar continuum within the proximity zone depends on various spectral features of the quasar on the red side of Ly$\alpha$ . If the continnum uncertainty of a specific quasar does not vary with wavelength over the short rest-frame range $1216 -1190 ~\rm \AA$, then the uncertainty in the peak-finding algorithm will be minimum; otherwise, the uncertainty analysis needs to be embedded into the continuum modeling, which requires Bayesian analysis using multi-parameter models \citep{hennawi2025}. We leave this for future studies.}

\section{Conclusions and Discussion}\label{sec:conclusions}

In this work, we conducted a proof-of-concept study to showcase how quasar proximity zones can be used to estimate the line-of-sight distance $\dlos$ between quasars in a projected quasar pair, thereby constraining their 3D separation.  We used synthetic quasar spectra to test a simple peak finding method for $\dlos$ estimation for several quasar pair scenarios and find the following:
\begin{itemize}
    \item For cases of small $\dsky\leq0.1$~pMpc: a simple peak finding method on the Gaussian smoothed transmission flux profile yields $\dlos$ estimates with a small error of $\approx 0.1$~\rm pMpc when the true $\dlos\gtrsim 2$~pMpc. While the peak-finding method cannot provide precise position estimates of the secondary quasar for smaller true $\dlos$, our method robustly constrains these cases to be $<2$~pMpc. (Figure~\ref{fig:predict_dlos_boxplot1})
    \item For cases of a larger $\dsky=1$~pMpc: a simple peak finding method on the Gaussian smoothed transmission flux profile yields $\dlos$ estimates with notably larger errors, where the contribution from the foreground quasar is relatively weaker.  The method yields an effective upper limit for distances $<4$~pMpc, and constrains the $\dlos$ with errors of $\approx1$~pMpc for true distances $\dlos>4$~pMpc. 
    (Figure~\ref{fig:boxplot_case3})
\end{itemize}

This study shows that for $\dsky \lesssim 0.1$ pMpc quasar pairs, proximity zone spectra can constrain the $\dlos$ within $<2$~pMpc and to better accuracy with larger $\dlos$.  This approach is an improvement from estimates that only rely on emission lines{, which have systematics errors of up to $\sim 6$ pMpc}. We also note that this proof-of-concept study utilizes only a simple single-peak finding method for line-of-sight distance estimation. Our results motivate further experiments to identify excess transmission (in particular, flux boost on small scales) due to the foreground quasar, which provides promising constraints of quasar pair configurations. More robust estimation would require careful analysis of small scale absorption features.  We caveat that our experiments assume negligible uncertainty in quasar continuum modeling. 
However, utilizing all absorption features on both large and small scales also makes the prediction routine less susceptible to such uncertainty, leveraging the robustness of small scale fluctuations to the quasar continuum.   

The relative improvement in accuracy of quasar pair distance estimation with such a simple method as ours motivates the exploration of more complex methods on proximity zone spectra, such as simulation-based inference, which might improve the prediction accuracy of parameters from 1-dimensional data vectors \citep{chen2023,li2024}.   We leave such experiments to future work.

\section*{Acknowledgments}
HC thanks the support of the Natural Sciences and Engineering Research Council of Canada (NSERC), funding reference number RGPIN-2025-04798 and DGECR-2025-00136, and by the Department of Science at the University of Alberta Augustana Campus. CA acknowledges support from the Leinweber Institute for Theoretical Physics at the University of Michigan.  JW acknowledges support from the Summer Undergraduate Research Fellowship program hosted by the Physics Department at the University of Michigan.


\bibliography{main}{}
\bibliographystyle{aasjournalv7}



\end{document}